\documentclass[11pt]{article}

\usepackage[normalem]{ulem}
\usepackage{amssymb}
\usepackage{amsmath}
\usepackage{amsthm}
\usepackage{setspace}
\usepackage{cite}
\usepackage{etoolbox}
\let\bbordermatrix\bordermatrix
\patchcmd{\bbordermatrix}{8.75}{4.75}{}{}
\patchcmd{\bbordermatrix}{\left(}{\left[}{}{}
\patchcmd{\bbordermatrix}{\right)}{\right]}{}{}

\newcommand{\NW}{\mbox{\small\rm NW}}
\newcommand{\IMM}{\mbox{\small\rm IMM}}
\newcommand{\VP}{\mbox{\small\rm VP}}
\newcommand{\VNP}{\mbox{\small\rm VNP}}
\renewcommand{\P}{\mbox{\small\rm P}}
\newcommand{\NP}{\mbox{\small\rm NP}}
\newcommand{\ex}{\mbox{\small\rm X}}
\newcommand{\hess}{\mbox{\small\rm H}}
\newcommand{\mathify}[1]{\mbox{\small\rm  #1}}
\newcommand{\dc}{\mbox{\small\rm dc}}
\newcommand{\F}{\mathbb{F}}
\newcommand{\LM}{\mathsf{LM}}
\newcommand{\depthfour}{\Sigma\Pi^{[O(\sqrt{n})]}\Sigma\Pi^{[\sqrt{n}]}}
\newcommand{\depthfourparam}{\Sigma\Pi^{[D]}\Sigma\Pi^{[t]}}
\usepackage{tikz}
\usetikzlibrary{matrix,positioning}

\usepackage[hmargin=2.5cm,vmargin=3cm]{geometry}
\newtheorem{definition}{Definition}
\newtheorem{corollary}{Corollary}
\newtheorem{prop}{Proposition}

\newtheorem{lemma}{Lemma}
\newtheorem{theorem}{Theorem}

\DeclareMathOperator{\rank}{\mbox{\small\rm rank}}
\DeclareMathOperator{\diag}{\mbox{\small\rm diag}}
\DeclareMathOperator{\perm}{\mbox{\small\rm perm}}

\title{Depth-4 Lower Bounds, Determinantal Complexity : A Unified Approach}
\author{Suryajith Chillara \qquad Partha Mukhopadhyay \\ \small{Chennai Mathematical Institute, India}\\\small{\{suryajith, partham\}@cmi.ac.in}}
\date{}

\begin{document}
\maketitle

\begin{abstract}
Tavenas has recently proved that any $n^{O(1)}$-variate and degree $n$ polynomial in $\VP$ can be computed by a depth-4 $\depthfour$ circuit of size $2^{O(\sqrt{n}\log n)}$\cite{tav2013}. So to prove $\VP\neq \VNP$, it is sufficient to show that an explicit polynomial $\in \VNP$ of degree $n$ requires $2^{\omega(\sqrt{n}\log n)}$ size depth-4 circuits. Soon after Tavenas's result, for two different explicit polynomials, depth-4 circuit size lower bounds of $2^{\Omega(\sqrt{n}\log n)}$ have been proved (\cite{kss2013} and \cite{flms2013}). In particular, using combinatorial design Kayal et al.\ \cite{kss2013} construct an explicit polynomial in $\VNP$ that requires depth-4 circuits of size $2^{\Omega(\sqrt{n}\log n)}$ and Fournier et al.\ show that the iterated matrix multiplication polynomial (which is in $\VP$) also requires $2^{\Omega(\sqrt{n}\log n)}$ size depth-4 circuits. 

In this paper, we identify a simple combinatorial property such that any polynomial $f$ that satisfies this property would achieve similar depth-4 circuit size lower bound. In particular, it does not matter whether $f$ is in $\VP$ or in $\VNP$. As a result, we get a simple unified lower bound analysis for the above mentioned polynomials.  

Another goal of this paper is to compare our current knowledge of the depth-4 circuit size lower bounds and the determinantal complexity lower bounds. Currently the best known determinantal complexity lower bound is $\Omega(n^2)$ for Permanent of $n\times n$ matrix (which is a $n^2$-variate and degree $n$ polynomial)\cite{ccl2008}. We prove that the determinantal complexity of the iterated matrix multiplication polynomial is $\Omega(dn)$ where $d$ is the number of matrices and $n$ is the dimension of the matrices. So for $d=n$, we get that the iterated matrix multiplication polynomial achieves the current best known lower bounds in both fronts: depth-4 circuit size and determinantal complexity. Our result also settles the determinantal complexity of the iterated matrix multiplication polynomial to $\Theta(dn)$.

To the best of our knowledge, a $\Theta(n)$ bound for the determinantal complexity for the iterated matrix multiplication polynomial was known only for a constant $d>1$\cite{jansen2011}.
\end{abstract}

\section{Introduction}
One of the main challenges in algebraic complexity theory is to separate $\VP$ from $\VNP$. This problem is well known as Valiant's hypothesis\cite{val1979}. This is an algebraic analog of the problem $\P$ vs $\NP$. Permanent polynomial characterizes the class 
$\VNP$ over fields of all characteristics except $2$ and the determinant polynomial characterizes the class $\VP$ with respect to quasi-polynomial projections.  
\begin{definition}
  The determinantal complexity of a polynomial $f$, over $n$ variables, is the minimum $m$ such that there are affine linear functions $A_{k,\ell}$, $1\leq k,\ell\leq m$ defined over the same set of variables and $f= \det((A_{k,\ell})_{1 \leq k,\ell \leq m})$. It is denoted by $\dc(f)$.  
\end{definition}
To resolve Valiant's hypothesis, proving $\dc(\perm_n) = n^{\omega(\log n)}$ is sufficient. Von zur Gathen \cite{von1986}  proved $\dc(\perm_n) \geq \sqrt{\frac{8}{7}}n$. Later Cai\cite{cai1990}, Babai and Seress \cite{von1987}, and Meshulam\cite{mesh1989} independently improved the lower bound to $\sqrt{2}n$. In 2004, Mignon and Ressayre\cite{mt2004} came up with a new idea of using second order derivatives and proved that $\dc(\perm_n) \geq \frac{n^2}{2}$ over the fields of characteristic zero. Subsequently, Cai et al.\cite{ccl2008} extended the result of Mignon and Ressayre to all fields of characteristic $\neq 2$.

For any polynomial $f$, Valiant\cite{val1979} proved that $\dc(f) \leq 2(F(f)+ 1)$ where $F(f)$ is the arithmetic formula complexity of $f$. Later, Nisan\cite{nisan1991} proved that $\dc(f) = O(B(f))$ where $B(f)$ is the the arithmetic branching program complexity of $f$. 

Another possible way to prove Valiant's hypothesis is to prove that the permanent polynomial can not be computed by any polynomial size arithmetic circuit. In 2008, Agrawal and Vinay proved that any arithmetic circuit of sub-exponential size can be depth reduced to a depth-4 circuit maintaining a nontrivial upper bound on the size \cite{av2008}. Subsequently, Koiran \cite{koi2012} and Tavenas \cite{tav2013} have come up with improved depth reductions (in terms of parameters). In particular, Tavenas proved that any $n^{O(1)}$-variate polynomial of degree $n$ in $\VP$ can also be computed by a $\depthfour$-circuit of top fan-in $2^{O(\sqrt{n}\log n)}$.  
   
In a recent breakthrough, Gupta et al. \cite{gkks2012} proved a $2^{\Omega(\sqrt{n})}$ lower bound for the size of the depth-$4$ circuits computing the determinant or the permanent polynomial using the method of shifted partial derivatives. Subsequently, Kayal et al. \cite{kss2013} improved the situation by proving a $2^{\Omega(\sqrt{n}\log n)}$ depth-4 $\Sigma\Pi^{[O(\sqrt{n})]}\Sigma\Pi^{[\sqrt{n}]}$-circuit size lower bound for an explicit polynomial in $\VNP$. 

More precisely, in \cite{kss2013} the following family of polynomials constructed from the combinatorial design of Nisan-Wigderson \cite{nw1994} were considered: 

\[
\NW(\ex)=\sum_{a(z)\in\mathbb{F}[z]} x_{1a(1)} x_{2a(2)}\ldots x_{na(n)}
\] 

where $a(z)$ runs over all univariate polynomials of degree $ < k$ for a suitable parameter $k=O(\sqrt{n})$ and $\mathbb{F}$ is a finite field of size $n$. Here we consider the natural identification of $\mathbb{F}$ with the set $\{1,2,\ldots,n\}$. Since the number of monomials in $\NW(\ex)$ is $n^{O(\sqrt{n})}$, the result from \cite{kss2013} gives a tight bound of $2^{\Theta{(\sqrt{n}\log n)}}$ for the depth-4 circuit complexity of $\NW(\ex)$.   

Although the combined implication of \cite{kss2013} and \cite{tav2013} looks very exciting from the perspective of lower bounds, a recent result by Fournier et al.\cite{flms2013} shows that such a lower bound is also obtained by the iterated matrix multiplication polynomial which is in $\VP$. The iterated matrix multiplication polynomial of $d$ generic $n\times n$ matrices $\ex^{(1)}, \ex^{(2)}, \ldots, \ex^{(d)}$ is the $(1,1)$th entry of the product of the matrices. More formally, let $\ex^{(1)}, \ex^{(2)}, \ldots, \ex^{(d)}$ be $d$ generic $n\times n$ matrices with disjoint set of variables and $x_{ij}^{(k)}$ be the variable in $\ex^{(k)}$ indexed by $(i,j)\in[n]\times[n]$.
Then the iterated matrix multiplication polynomial (denoted by $\IMM_{n,d}$) is defined as follows.  

\[
 \IMM_{n,d}(\ex) = \sum_{i_1, i_2, \ldots, i_{d-1} \in [n]}x_{1i_1}^{(1)}x_{i_1i_2}^{(2)}\dots x_{i_{(d-2)}i_{(d-1)}}^{(d-1)}x_{i_{(d-1)}1}^{(d)}  
\]

Notice that $\IMM_{n,d}(\ex)$ is a $n^2(d-2) + 2n$-variate polynomial of degree $d$. The result from \cite{flms2013} is also tight since $\IMM_{n,d}\in\VP$ and very importantly their result proves the optimality of the depth reduction of Tavenas \cite{tav2013}.

One of the main motivations of our study comes from this tantalizing fact that two seemingly different polynomials $\NW(\ex)\in\VNP$ and $\IMM_{n,d}(\ex)\in \VP$ behave very  similarly as far as the $2^{\Omega(\sqrt{n}\log n)}$-size lower bound for depth-4 circuits are concerned. In this paper, we seek a conceptual reason for this behaviour. We identify a simple combinatorial property such that any polynomial that satisfies it, would require $2^{\Omega(\sqrt{n}\log n)}$-size depth-4 arithmetic circuits. We call it \emph{Leading Monomial Distance Property}. In particular, it does not matter whether the polynomial is easy (i.e. in $\VP$) or hard (i.e. the polynomial is in $\VNP$ but not known to be in $\VP$). As a result of this abstraction we give a simple \emph{unified} analysis of the depth-4 circuit size lower bounds for $\NW(\ex)$ and $\IMM_{n,d}(\ex)$. 

To define the Leading Monomial Distance Property, we first define the notion of distance between two monomials.
\begin{definition}
  Let $m_1, m_2$ be two monomials over a set of variables. Let $S_1$ and $S_2$ be the (multi)-sets of variables corresponding to the monomials $m_1$ and $m_2$ respectively. The distance $\Delta(m_1, m_2)$ between the monomials $m_1$ and $m_2$ is the $\min\{|S_1|-|S_1\cap S_2|, |S_2|-|S_1\cap S_2|\}$ where the cardinalities are the order of the multi-sets. 
\end{definition}

For example, let $m_1 = x_1^2x_2x_3^2x_4$ and $m_2 = x_1x_2^2x_3x_5x_6$. Then $S_1 = \{x_1, x_1, x_2, x_3, x_3, x_4\}$, $S_2 = \{x_1, x_2, x_2, x_3, x_5, x_6\}$ and $\Delta(m_1, m_2) = 3$. 

We say that a $n^{O(1)}$-variate and $n$-degree polynomial has the Leading Monomial Distance Property, if the leading monomials of a \emph{large subset} ($\approx n^{\sqrt{n}}$) of its span of the derivatives (of order $\approx \sqrt{n}$) have \emph{good pair-wise distance}. Leading monomials are defined by defining a suitable order on the set of variables. We denote the leading monomial of a polynomial $f(\ex)$ by $\LM(f)$. More formally, we prove the following theorem in Section~\ref{unifiedsec}.

\begin{theorem}\label{depth4intro}
Let $f(\ex)$ be a $n^{O(1)}$-variate polynomial of degree $n$. Let there be $s\geq n^{\delta k}$ ($\delta$ is some constant $>0$) different polynomials in $\langle\partial^{=k}(f)\rangle$ for $k=\epsilon\sqrt{n}$ (where $0<\epsilon<1$ is any constant) such that any two of their leading monomials have pair-wise distance of at least $d\geq \frac{n}{c}$ for a constant $c>1$. Then any depth-4 $\depthfour$ circuit that computes $f(\ex)$ must be of size $e^{\Omega(\sqrt{n}\ln n)}$.
\end{theorem}

Another motivation of this work is to find a connection between our current knowledge of the determinantal complexity lower bounds and the depth-4 circuit size lower bounds. The best known determinantal complexity lower bound for a $n^{O(1)}$-variate and $n$ degree (Permanent) polynomial is $\Omega(n^2)$. Here we ask the following question: can we give an example of an explicit $n^{O(1)}$-variate degree $n$ polynomial in $\VNP$ for which the determinantal complexity is $\Omega(n^2)$ and the the depth-4 complexity is $2^{\Omega(\sqrt{n}\log n)}$ ? We settle this problem by showing a $\Omega(n^2)$ lower bound for $\dc(\IMM_{n,n}(\ex))$ which is a $O(n^3)$-variate and $n$-degree polynomial. In particular, we prove the following theorem. 

\begin{theorem}
  \label{detcompIMM}
  For any integers $n$ and $d>1$, the determinantal complexity of the iterated matrix multiplication polynomial $\IMM_{n,d}$ is $\Omega(dn)$. 
\end{theorem}

Since $\IMM_{n,d}(\ex)$ has an algebraic branching program of size $O(dn)$ \cite{nisan1991}, from the above theorem it follows that $\dc(\IMM_{n,d}(\ex)) = \Theta(dn)$. This improves upon the earlier bound of $\Theta(n)$ for the determinantal complexity of the iterated matrix multiplication polynomial for a constant $d>1$ \cite{jansen2011}. Similar to the approach of \cite{ccl2008} and \cite{mt2004}, we also use the the rank of Hessian matrix as our main technical tool.

\section{Organization}
\label{orgsec}
In Section~\ref{prelim}, we state some results from \cite{gkks2012}. In Section~\ref{unifiedsec}, we do a unified analysis of the depth-4 lower bound results of \cite{kss2013} and \cite{flms2013}. We prove the determinantal complexity lower bound of $\IMM_{n,d}(\ex)$ in Section~\ref{detcompsec}. 

\section{Preliminaries}
\label{prelim}
 The following beautiful lemma (from \cite{gkks2012}) is the key to the asymptotic estimates required for the lower bound analyses. 
\begin{lemma}
  \label{bin}
  Let $a(n), f(n), g(n) : \mathbb{Z}_{\geq 0} \rightarrow \mathbb{Z}_{\geq 0}$ be the integer valued functions such that $(f+g) = o(a)$, then
  \begin{align*}
    \ln \frac{(a+f)!}{(a-g)!} = (f+g)\ln a \pm O\left(\frac{(f+g)^2}{a}\right)
  \end{align*}
\end{lemma}

In this paper, whenever we apply this lemma, $(f+g)^2$ will be $o(a)$. So, we will not worry about the error term (which will be asymptotically zero) generated by this estimate.

Building on the results of \cite{av2008} and \cite{koi2012}, Tavenas\cite{tav2013} proved that any $n^{O(1)}$-variate polynomial of degree $n$ in $\VP$ can be computed by a $\Sigma\Pi^{[O(\sqrt{n})]}\Sigma\Pi^{[\sqrt{n}]}$ circuit of top fan-in $2^{O(\sqrt{n}\log n)}$. 

In \cite{gkks2012}, the authors introduced the method of \emph{shifted partial derivatives} which is the key tool in the recent depth-4 lower bound results. They also gave an elegant way of upper bounding the dimension of the shifted partial derivative space of a polynomial computed by a $\depthfour$ circuit.

For a monomial $\mathbf{x}^{\mathbf{i}} = x_1^{i_1}x_2^{i_2}\dots x_n^{i_n}$, let $\partial^{\mathbf{i}}f$ be the partial derivative of $f$ with respect to the monomial $\mathbf{x}^{\mathbf{i}}$. We recall the following definition of shifted derivatives from \cite{gkks2012}.
\begin{definition}
  \label{shiftedPartialDerDefn}
  Let $f(\ex)\in \mathbb{F}[\ex]$ be a multivariate polynomial. The span of the $\ell$-shifted $k$-th order derivatives of $f$, denoted by $\langle\partial^{=k}f\rangle_{\leq \ell}$, is defined as
  \begin{align*}
    \langle\partial^{=k}f\rangle_{\leq \ell} = \F\mbox{-span} \{\mathbf{x}^{\mathbf{i}}\cdot(\partial^{\mathbf{j}}f) : \mathbf{i}, \mathbf{j} \in \mathbb{Z}^n_{\geq 0} \mbox{ with } |\mathbf{i}| \leq \ell \mbox{ and } |\mathbf{j}| = k \}
  \end{align*}
  We denote by $\dim(\langle\partial^{=k}f\rangle_{\leq \ell})$ the dimension of the vector space $\langle\partial^{=k}f\rangle_{\leq \ell}$.
\end{definition}

The following proposition follows from the standard Gaussian elimination technique.
\begin{prop}[Corollary 13, \cite{gkks2012}]
\label{leadingmonomialcount}
  For any multivariate polynomial $f(\ex)\in \mathbb{F}[\ex]$,
  \begin{align*}
    \dim(\langle\partial^{=k}f\rangle_{\leq \ell}) \geq \#\{\mathbf{x}^{\mathbf{i}}\cdot\LM(g) : \mathbf{i}, \mathbf{j} \in \mathbb{Z}^n_{\geq 0} \mbox{ with } |\mathbf{i}| \leq \ell \mbox{ and } |\mathbf{j}| = k \mbox{ and } g\in \mathbb{F}\mbox{-span}\{\partial^{\mathbf{i}}f\}\}
  \end{align*}

\end{prop}
In \cite{gkks2012}, the following upper bound on the dimension of the shifted partial derivative space for polynomials computed by $\depthfourparam$ circuits was shown. 
\begin{lemma}[Corollary 10, \cite{gkks2012}]
  \label{GKKSupperbound}
  If $C = \sum_{i=1}^{s'}Q_{i1}Q_{i2}\dots Q_{iD}$ where each $Q_{ij}\in\mathbb{F}[\ex_N]$ is a polynomial of degree bounded by $t$, then for any $k\leq D$
  \begin{align*}
    \dim(\langle \partial^{=k}(C)\rangle_{\leq \ell}) \leq s'{D+k\choose k}{N+\ell+k(t-1)\choose N}
  \end{align*}
\end{lemma}

\section{Unified analysis of depth-4 lower bounds}
\label{unifiedsec}
In this section we first prove a simple combinatorial lemma which we believe is the crux of the best known depth-4 lower bound results. In fact, the lower bounds on the size of $\depthfour$ circuits computing the polynomials $\NW(\ex)$ and $\IMM_{n,d}(\ex)$ follow easily from this lemma by suitable setting of parameters. 
 
\begin{lemma}
  \label{distanceLemma}
  Let $m_1, m_2, \dots, m_s$ be the monomials over $N$ variables such that $\Delta(m_i, m_j)\geq d$. Each monomial $m_i$ is extended by padding monomials of length at most $\ell$ over $N$ variables. Let $B={N+\ell\choose N}$. Then the total number of monomials generated by the extension is at least $\left(sB - s^2{N+\ell-d\choose N}\right)$.
\end{lemma}
\begin{proof}
  Let $B_i$ be the set of monomials constructed by extending the monomial $m_i$. It is easy to see that $|B_i| = {N+\ell\choose N}$. We would like to estimate $|\cup_i B_i|$. From the principle of inclusion and exclusion, we get
  \begin{align*}
    |\cup_{i = 1}^s B_i| &\geq \sum_{i\in [s]} |B_i| - \sum_{i,j\in [s], i\neq j} |B_i\cap B_j|
  \end{align*}

  Now we estimate the upper bound for $|B_i\cap B_j|$ such that $i\neq j$. Consider the monomials $M_i$ and $M_j$ in $B_i$ and $B_j$ respectively. For $M_i$ and $M_j$ to match, $M_i$ should contain at least $d$ variables from $m_j$ and similarly $M_j$ should contain at least $d$ variables from $m_i$. The rest of the at most $(\ell-d)$ degree monomials should be identical in $M_i$ and $M_j$. The number of such monomials over $N$ variables is at most ${N+\ell-d\choose N}$. Thus,
  \begin{align*}
    |B_i \cap B_j| \leq {N+\ell-d\choose N}
  \end{align*}

  Then the total number of monomials after extension is lower bounded as follows.
  \begin{align*}
    |\cup_{i = 1}^s B_i| &\geq sB - s^2{N+\ell-d\choose N} = sB\left(1-\frac{s}{B}{N+\ell-d\choose N}\right)
  \end{align*} 
\end{proof}

To get a good lower bound for $|\cup_{i=1}^s B_i|$, we need to upper bound $\frac{s}{B}{N+\ell-d\choose N}$. Let us bound it by an inverse polynomial in $N$ by suitably choosing $\ell$: $\frac{s{N+\ell-d\choose N}}{{N+\ell\choose N}} \leq\frac{1}{p(N)}$ where $p(N)$ is a polynomial in $N$. 

After simplification, we get $ s \frac{(N + \ell -d)!}{(N + \ell)!} \frac{\ell!}{(\ell-d)!}\leq\frac{1}{p(N)}$. We will use Lemma \ref{bin} to tightly estimate the subsequent computations. In particular, we always choose parameter $\ell$ such that $d^2=o(N +\ell)$. This also shows that the error term given by Lemma \ref{bin} is always asymptotically zero and we will not worry about it. 

We now apply Lemma~\ref{bin} to derive the following: $s \left(\frac{\ell}{N + \ell}\right)^d\leq \frac{1}{p(N)}$ or equivalently $s\left(\frac{1}{1+\frac{N}{\ell}}\right)^d\leq\frac{1}{p(N)}$. We use the inequality $1 + x > e^{x/2}$ for $0< x < 1$ to lower bound $\left(1+\frac{N}{\ell}\right)^d$ by $e^{\frac{Nd}{2l}}$. Thus, it is enough to choose $\ell$ in a way that $s\cdot p(N)\leq e^{\frac{N d}{2\ell}}$ or equivalently $\ell \leq \frac{Nd}{2\ln (s\cdot p(N))}$.

Let $f$ be a polynomial such that there are at least $s$ polynomials in $\langle\partial^{=k}(f)\rangle$ (for a suitable value of $k$) and any two of their leading monomials have distance of at least $d$. Then for the above choice of $\ell$ and from Proposition~\ref{leadingmonomialcount}, we know that the dimension of $\langle\partial^{=k}f\rangle_{\leq \ell} \geq (1-\frac{1}{p(N)}) ~s{{N +\ell}\choose{N}}$.

Combining this with Lemma~\ref{GKKSupperbound}, we get the following :  $s' \geq \left(1-\frac{1}{p(N)}\right)~s~\frac{{N+l\choose N}}{{D+k\choose k}{N+l+k(t-1)\choose N}}$. Suppose we choose $\ell$ such that $(kt-k)^2=o(\ell)$. Then by applying Lemma~\ref{bin} we can easily show that $s'\geq  \frac{s\left(1-\frac{1}{p(N)}\right)}{{D+k\choose k} {(1+\frac{N}{l})^{(kt-k)}}}\geq \frac{s\left(1-\frac{1}{p(N)}\right)}{{D+k\choose k} e^{\frac{N}{\ell}(kt-k)}}$. 

We get the following Theorem from the above discussion. 

\begin{theorem}\label{depth4}
  Let $f(\ex)$ be a $n^{O(1)}$-variate polynomial of degree $n$. Let there be $s\geq n^{\delta k}$ ($\delta$ is some constant $>0$) different polynomials in $\langle\partial^{=k}(f)\rangle$ for $k=\epsilon\sqrt{n}$ (where $0<\epsilon<1$ is any constant) such that any two of their leading monomials have distance of at least $d\geq \frac{n}{c}$ for a constant $c>1$. Then any depth-4 $\depthfour$ circuit that computes $f(\ex)$ must be of size $s'\geq e^{\Omega(\sqrt{n}\ln n)}$.
\end{theorem}  

\begin{proof}
All we need to do is to choose $\ell$ appropriately so that we get the desired lower bound. Let $N$ be the number of variables in $f$. Then from the upper bound estimate of $\ell$, we get that $\ell\leq \frac{N(n/c)}{2(\delta k \ln n + 2\ln n)}$ by fixing the inverse polynomial to $\frac{1}{n^2}$. So it is enough to choose $\ell$ such that $\ell\leq\frac{Nn}{4c\delta k\ln n}=\frac{N\sqrt{n}}{4c\delta\epsilon\ln n}$.

From the lower bound arguments mentioned above, we know that $s'\geq \frac{(1-\frac{1}{n^2})n^{\delta k}}{{{D+k}\choose{k}}e^{\frac{N}{\ell}kt}}$. Since $D=c'\sqrt{n}$ for some $c' > 0$ and $k=\epsilon\sqrt{n}$, ${{D+k}\choose{k}}=e^{O(\sqrt{n})}$ by Shannon's entropy estimate for binomial coefficient. To get the required lower bound it is enough to choose $\ell$ such that $\frac{N k t}{\ell}<\mu \delta k\ln n$ for some $0<\mu<1$. Since $t\leq \sqrt{n}$, it is enough to take $\ell > \frac{N\sqrt{n}}{\mu\delta\ln n}$. 

By comparing the lower and upper bounds of $\ell$, we get that $\epsilon < \frac{\mu}{4c}$. Once we fix $\mu$, we need to choose $\epsilon < \min\{c', \frac{\mu}{4c}\}$.      
\end{proof}

In the next section, we show that the lower bounds on the size of $\depthfour$ circuits computing $\NW(\ex)$ and $\IMM_{n,n}(\ex)$ can be obtained by simply applying the Theorem~\ref{depth4}. Moreover, it shows that the lower bound arguments of $\IMM_{n,n}(\ex)$ are essentially same as the lower bound arguments of $\NW(\ex)$.

\subsection{Lower bounds on the size of depth-4 circuits computing $\NW(\ex)$ and $\IMM_{n,d}(\ex)$}
\label{lower-bound}

Now we derive the depth-4 circuit size lower bound for $\NW(\ex)$ polynomial by a simple application of Theorem~\ref{depth4}. 
 
\begin{corollary}\label{NWcor}
  Any depth-4 $\depthfour$ circuit computing the $\NW(\ex)$ polynomial over $n^2$ variables would be of size $2^{\Omega(\sqrt{n}\log n)}$. 
\end{corollary}

\begin{proof}
Recall that 
$\NW(\ex)=\sum_{a(z)\in \mathbb{F}[z]}x_{1 a(1)}x_{2 a(2)}\ldots x_{n a(n)}$ where $\mathbb{F}$ is a finite field of size $n$ and $a(z)$ are the polynomials of degree $\leq k-1$. Notice that any two monomials can intersect in at most $k-1$ variables. Here we fix an ordering on the variables: $x_{11}\succ x_{12} \succ \dots \succ x_{nn}$.

We differentiate the polynomial $\NW(\ex)$ with respect to the first $k=\epsilon\sqrt{n}$ variables of each monomial. After differentiation, we get $n^k$ monomials of length $(n-k)$ each. Since they are constructed from the image of univariate polynomials of degree at most $(k-1)$, the distance $d$ between any two monomials $\geq n-2k > n/2$. So to get the required lower bound we apply Theorem~\ref{depth4} with $\delta=1$ and $c=2$. The parameter $\epsilon$ will get fixed from Theorem~\ref{depth4}.

\end{proof}

Next we derive the lower bound on the size of the depth-4 circuit computing $\IMM_{n,n}(\ex)$.
  
\begin{corollary}\label{IMMcor}
  Any depth-4 $\depthfour$ circuit computing the $\IMM_{n,n}(\ex)$ polynomial would be of size $2^{\Omega(\sqrt{n}\log n)}$. 
\end{corollary}

\begin{proof}
Recall that $\IMM_{n,n}(\ex) = \sum_{i_1, i_2, \dots, i_{n-1} \in [n]}x_{1i_1}^{(1)}x_{i_1i_2}^{(2)}\ldots x_{i_{(n-2)}i_{(n-1)}}^{(n-1)}x_{i_{(n-1)}1}^{(n)}$. It is a polynomial over $(n-2)n^2 + 2n$ variables. Next we fix the following lexicographic ordering on the variables of the set of matrices $\{\ex^{(1)},\ex^{(2)},\ldots,\ex^{(n)}\}$ as follows: $\ex^{(1)} \succ \ex^{(2)} \succ \ex^{(3)} \succ \ldots \succ \ex^{(n)}$ and in any $\ex^{(i)}$ the ordering is $x^{(i)}_{11}\succ x^{(i)}_{12}\succ \ldots \succ x^{(i)}_{1n}\succ\ldots \succ x^{(i)}_{n1}\ldots\succ x^{(i)}_{nn}$.

Choose a prime $p$ such that $\frac{n}{2} \leq p \leq n$. Consider the set of univariate polynomials $a(z)\in \mathbb{F}_p[z]$ of degree at most $(k-1)$ for $ k = \epsilon \sqrt{n}$ where $\epsilon$ is a small constant to be fixed later in the analysis.  

Consider a set of $2k$ of those matrices such that they are $n/4k$ distance apart : $\ex^{(2)}, \ex^{(3 + \frac{n}{4k})}, \ldots, \ex^{(2k+1 + \frac{(2k-1)n}{4k})}$. Clearly $2k + 1 + \frac{(2k-1)n}{4k}<n$.  
For each univariate polynomial $a$ of degree at most $(k-1)$, define a set $S_a = \{x^{(2)}_{1,a(1)}, x^{(3+\frac{n}{4k})}_{2,a(2)}, \dots, x^{(2k+1 + \frac{(2k-1)n}{4k})}_{2k,a(2k)}\}$. Number of such sets is at least $\left(\frac{n}{2}\right)^k$ and $|S_a \cap S_b|< k$ for $a\neq b$. Now we consider a polynomial $f(\ex)$ which is a restriction of the polynomial $\IMM_{n,n}(\ex)$. By restriction we simply mean that a few variables of $\IMM_{n,n}(\ex)$ are fixed to some elements from the field and the rest of the variables are left untouched. We define the restriction as follows. 
\[
x_{ij}^{(q)}=0 ~\mbox{if}~ r + \frac{(r-2)n}{4k} < q < (r+1) + 
\frac{(r-1)n}{4k}-1 
~\mbox{for some} ~2\leq r \leq 2k ~\mbox{and}~ i \neq j. 
\]

The rest of the variables are left untouched.     

Next we differentiate the polynomial $f(\ex)$ with respect to the sets of variables $S_a$ indexed by the polynomials $a(z)\in\mathbb{F}[z]$. Consider the leading monomial of the derivatives with respect to the sets $S_a$ for all $a(z) \in \mathbb{F}[z]$. Since $|S_a\cap S_b|< k$, it is straightforward to observe that the distance between any two leading monomials is at least $k\cdot\frac{n}{4k}=\frac{n}{4}$. The intuitive justification is that whenever there is a difference in $S_a$ and $S_b$, that difference can be stretched to a distance $\frac{n}{4k}$ because of the restriction that eliminates non diagonal entries.   

Now we prove the lower bound for the polynomial $f(\ex)$ by applying Theorem~\ref{depth4}. Clearly $s\geq(n/2)^k>n^{\frac{1}{4}(2k)}$. So we set the parameter $\delta=1/4$ and $c=4$. The parameter $\epsilon$ will get fixed from Theorem~\ref{depth4}. Since $f(\ex)$ is a restriction of $\IMM_{n,n}(\ex)$, any lower bound for $f(\ex)$ is a lower bound for $\IMM_{n,n}(\ex)$ too : If $\IMM_{n,n}(\ex)$ has a $2^{o(\sqrt{n}\log n)}$ sized $\depthfour$ circuit, then we get a $2^{o(\sqrt{n}\log n)}$ sized $\depthfour$ circuit for $f(\ex)$ by fixing the variables according to the restriction.
\end{proof}

\section{Determinantal complexity of $\IMM_{n,d}$}
\label{detcompsec}
We start by recalling a few facts from \cite{ccl2008}. Let $A_{k,\ell}(\ex)$, $1\leq k,\ell\leq m$ be the affine linear functions over $\mathbb{F}[\ex]$ such that the following is true.
\begin{align*}
  \IMM_{n,d}(\ex) = \det((A_{k,\ell}(\ex))_{1\leq k,\ell\leq m})
\end{align*}
 Consider a point $\ex_0\in \mathbb{F}^{n^2d}$ such that $\IMM_{n,d}(\ex_0)=0$. The affine linear functions $A_{k,\ell}(\ex)$ can be expressed as $L_{k,\ell}(\ex-\ex_0) + y_{k,\ell}$ where $L_{k,\ell}$ is a linear form and $y_{k,\ell}$ is a constant from the field. Thus, $(A_{k,\ell}(\ex))_{1\leq k,\ell\leq m} = (L_{k,\ell}(\ex-\ex_0))_{1\leq k,\ell\leq m} + \mathify{Y}_0$. If $\IMM_{n,d}(\ex_0)=0$ then $\det(\mathify{Y}_0)=0$. Let $\mathify{C}$ and $\mathify{D}$ be two non-singular matrices such that $\mathify{C}\mathify{Y}_0\mathify{D}$ is a diagonal matrix.

\begin{align*}
  \mathify{C}\mathify{Y}_0\mathify{D} =
  \begin{pmatrix}
    0& 0\\
    0& \mathify{I}_s\\
  \end{pmatrix}
\end{align*}

Since $\det(\mathify{Y}_0)=0$, $s<m$. From the previous works \cite{von1987}, \cite{cai1990}, \cite{mt2004}, and \cite{ccl2008}, it is enough to assume that $s=m-1$. Since the first row and the first column of $\mathify{C}\mathify{Y}_0\mathify{D}$ are zero, we may multiply $\mathify{C}\mathify{Y}_0\mathify{D}$ by $\diag(\det(\mathify{C})^{-1},1,\dots,1)$ and $\diag(\det(\mathify{D})^{-1},1,\dots,1)$ on the left and the right side. Without loss of generality, we may assume that $\det(\mathify{C})=\det(\mathify{D})=1$. By multiplying with $\mathify{C}$ and $\mathify{D}$ on the left and the right and suitably renaming $(L_{k,\ell}(\ex-\ex_0))_{1\leq k,\ell\leq m}$ and $\mathify{Y}_0$ we get
\begin{align*}
  \IMM_{n,d}(\ex) = \det((L_{k,\ell}(\ex-\ex_0)_{1\leq k,\ell\leq m} + \mathify{Y}_0))
\end{align*}
where $\mathify{Y}_0 = \diag(0,1,\dots,1)$. 

We use $\hess_{\IMM_{n,d}}(\ex)$ to denote the Hessian matrix of the iterated matrix multiplication and is defined as follows.
\begin{align*}
  \hess_{\IMM_{n,d}}(\ex) &= (H_{s;ij,t;k\ell}(\ex))_{1\leq i,j\leq n, 1\leq s,t \leq d }\\
  H_{s;ij,t;k\ell}(\ex) &= \frac{\partial^2\IMM_{n,d}(\ex)}{\partial x_{ij}^{(s)}\partial x_{k\ell}^{(t)}}
\end{align*}
where $x_{ij}^{(s)}$ and $x_{k\ell}^{(t)}$  denote the $(i,j)$th and $(k,\ell)$th entries of the variable sets $\ex^{(s)}$ and $\ex^{(t)}$ respectively. 

By taking second order derivatives and evaluating the Hessian matrices of $\IMM_{n,d}(\ex)$ and $\det((A_{k,\ell}(\ex))_{1\leq k,\ell\leq m})$ at $\ex_0$, we obtain $\hess_{\IMM_{n,d}}(\ex_0) = \mathify{L}\hess_{\det}(\mathify{Y}_0)\mathify{L}^{T}$ where $\mathify{L}$ is a $n^2d\times m^2$ matrix with entries from the field. It follows that $\rank(\hess_{\IMM_{n,d}}(\ex_0)) \leq \rank(\hess_{\det}(\mathify{Y}_0))$. We give an explicit construction of a point $\ex_0 \in \mathbb{F}^{n^2d}$ such that $\IMM_{n,d}(\ex_0) = 0$ and $\rank(\hess_{\IMM_{n,d}}(\ex_0)) \geq d(n-1)$.

\subsection{Upper bound for the rank of $\hess_{\det}(\mathify{Y}_0)$}
\label{upperbound}
This analysis is the same as the one given in \cite{ccl2008} and \cite{mt2004} and we briefly recall it here for the sake of completeness. The second order derivative of $\det(\mathify{Y})$ with respect to the variables $y_{ij}$ and $y_{k\ell}$ eliminates the rows $\{i,k\}$ and the columns $\{j,\ell\}$. Considering the form of $\mathify{Y}_0$, the non-zero entries in $\hess_{\det}(\mathify{Y}_0)$ are obtained only if $1\in\{i,k\}$ and $1\in\{j,\ell\}$ and thus $(ij,k\ell)$ are of the form $(11,tt)$ or $(t1,1t)$ or $(1t,t1)$ for any $t>1$. This gives $\rank(\hess_{\det}(\mathify{Y}_0)) = O(m)$.

\subsection{Lower bound for the rank of $\hess_{\IMM_{n,d}}(\ex_0)$}
\label{lowerbound}
In this section, we prove Theorem~\ref{detcompIMM}. In particular, we give a polynomial time algorithm to construct a point $\ex_0$ explicitly such that $\IMM_{n,d}(\ex_0) = 0$ and $\rank(\hess(\IMM_{n,d}(\ex_0)))\geq d(n-1)$. From the discussion in Section~\ref{upperbound} and the upper bound for $\dc(\IMM_{n,d}(\ex))$ from \cite{nisan1991}, it is clear that such a construction is sufficient to prove Theorem~\ref{detcompIMM}. 

\begin{theorem}
\label{immLowerBound}
  For any integers $n,d>1$, there is a point $\ex_0 \in \mathbb{F}^{n^2d}$ such that $\IMM_{n,d}(\ex_0) =0 $ and $\rank(\hess(\IMM_{n,d}(\ex_0)))\geq d(n-1)$. Moreover, the point $\ex_0$ can be constructed explicitly in polynomial time.
\end{theorem}
\begin{proof}
  We prove the theorem by induction on $d$. For the purpose of induction, we maintain that the entries indexed by the indices $(1,2), (1,3), \dots, (1,n)$ of the matrix obtained after multiplying the first $(d-1)$ matrices are not all zero at $\ex_0$.

  We first prove the base case for $d=2$. The corresponding polynomial is $\IMM_{n,2}(\ex) = \displaystyle\sum_{i=1}^nx_{1i}^{(1)}x_{i1}^{(2)}$. It is easy to observe that the rank of the Hessian matrix is $2n > 2(n-1)$ at any point since each non-zero entry of the Hessian matrix is $1$ and the structure of the Hessian matrix is the following: 
  
  \begin{align*}
    \hess_{\IMM_{n,2}}(\ex)=
    \begin{bmatrix}
      0& B_{12}\\
      B_{21}& 0
    \end{bmatrix}
  \end{align*}
  where the only non-zero rows of $B_{12}$ are shown in the figure below and $B_{21} = B_{12}^{T}$. 
  \begin{align*}
    \bbordermatrix
    {
      \text{}& x^{(2)}_{11}& x^{(2)}_{12}& \dotsm& x^{(2)}_{1n}& x^{(2)}_{21}& x^{(2)}_{22}& \dotsm& x^{(2)}_{2n}& \dotsm& x^{(2)}_{n1}& x^{(2)}_{n2}& \dotsm& x^{(2)}_{nn}\cr
      x^{(1)}_{11}& {1}& 0& \dotsm& 0& 0& 0& \dotsm& 0& \dotsm& 0& 0& \dotsm& 0\cr
      x^{(1)}_{12}& {0}& 0& \dotsm& 0& 1& 0& \dotsm& 0& \dotsm& 0& 0& \dotsm& 0\cr
      \vdots& \dotsm& \dotsm& \dotsm& \dotsm& \dotsm& \dotsm& \dotsm& \dotsm& \dotsm& \dotsm& \dotsm& \dotsm& \dotsm&\cr
      x^{(1)}_{1n}& {0}& 0& \dotsm& 0& 0& 0& \dotsm& 0& \dotsm& 1& 0& \dotsm& 0\cr
    }
  \end{align*}
  We set the values of the variables as follows: $x_{11}^{(1)} =0$, $x_{11}^{(2)} =1$, $x_{21}^{(2)}=x_{31}^{(2)}=\dots=x_{n1}^{(2)}=0$ and $x_{12}^{(1)}, x_{13}^{(1)},\dots,x_{1n}^{(1)}$ arbitrarily but not all zero. The point thus obtained (say $\ex_0$) is clearly a zero of the polynomial $\IMM_{n,2}(\ex)$.

  Now consider the case where the first $d$ matrices are multiplied. The final polynomial could be represented as follows.
  \begin{align*}
    \IMM_{n,d}(\ex) = \sum_{i_1, i_2, \dots, i_{d-1} \in [n]}x_{1i_1}^{(1)}x_{i_1i_2}^{(2)}\dots x_{i_{(d-2)}i_{(d-1)}}^{(d-1)}x_{i_{(d-1)}1}^{(d)}  
  \end{align*}
  For induction hypothesis, assume that the statement of the theorem is true for the case where the number of matrices being multiplied is $\leq d$. Consider the polynomial $\IMM_{n,(d+1)}(\ex)$. 
  \begin{align*}
    \IMM_{n,(d+1)}(\ex) = \sum_{i_1, i_2, \dots, i_{d-1}, i_d \in [n]}x_{1i_1}^{(1)}x_{i_1i_2}^{(2)}\dots x_{i_{(d-2)}i_{(d-1)}}^{(d-1)}x_{i_{(d-1)}i_d}^{(d)}x_{i_{d}1}^{(d+1)}
  \end{align*}
  Let the matrix obtained after multiplying the first $d$ matrices be the following.
  \begin{align*}
    \begin{bmatrix}
      P_{11}(\ex)& P_{12}(\ex)& \dotsm& P_{1n}(\ex)\\
      P_{21}(\ex)& P_{22}(\ex)& \dotsm& P_{2n}(\ex)\\
      \vdots& \vdots& \ddots& \vdots\\
      P_{n1}(\ex)& P_{n2}(\ex)& \dotsm& P_{nn}(\ex)
    \end{bmatrix}
  \end{align*} where
  \begin{align*}
    P_{k\ell}(\ex) &= \sum_{i_1, i_2, \dots, i_{d-1} \in [n]}x_{ki_1}^{(1)}x_{i_1i_2}^{(2)}\dots x_{i_{(d-2)}i_{(d-1)}}^{(d-1)}x_{i_{(d-1)}\ell}^{(d)} \\
  \end{align*}
  Thus, we have the following expression.
  \begin{align*}
    \IMM_{n,(d+1)}(\ex) = P_{11}(\ex)x_{11}^{(d+1)}+P_{12}(\ex)x_{21}^{(d+1)}+\dots+P_{1n}(\ex)x_{n1}^{(d+1)}
  \end{align*}

  Now consider the Hessian matrix $\hess_{\IMM_{n, d+1}}(\ex)$ which is a $(d+1)n^2\times (d+1)n^2$ sized matrix.

  \begin{align*}
    \hess_{\IMM_{n,d+1}}(\ex) =
    \begin{bmatrix}
      0& B_{1,2}& B_{1,3}& B_{1,4}& \dotsm& B_{1,(d+1)}\\
      B_{2,1}& 0& B_{2,3}& B_{2,4}& \dotsm& B_{2,(d+1)}\\
      B_{3,1}& B_{3,2}& 0& B_{3,4}& \dotsm& B_{3,(d+1)}\\
      \vdots& \vdots& \vdots& \ddots& \vdots& \vdots \\
      \vdots& \vdots& \vdots& \vdots& \ddots& \vdots \\
      B_{(d+1),1}& B_{(d+1),2}& \dotsm& \dotsm& B_{(d+1),d}& 0
    \end{bmatrix}
  \end{align*}

  Each $B_{i,j}$ is a block of size $n^2\times n^2$ which is indexed by the variables from the matrices $M^{(i)}$ and $M^{(j)}$ with the corresponding variable sets $\ex^{(i)}$ and $\ex^{(j)}$. Consider the block $B_{(d+1),d}$ which is indexed by the variable sets $\ex^{(d+1)}$ and $\ex^{(d)}$. The only non-zero rows in $B_{(d+1),d}$ are indexed by the variables $x^{(d+1)}_{11}, x^{(d+1)}_{21}, \dots, x^{(d+1)}_{n1}$. The potential non-zero entries for the row $x^{(d+1)}_{11}$ are indexed by the columns $x^{(d)}_{11}, x^{(d)}_{21}, \dots, x^{(d)}_{n1}$. Similarly the potential non-zero entries for the row $x^{(d+1)}_{21}$ are indexed by the columns $x^{(d)}_{12}, x^{(d)}_{22}, \dots, x^{(d)}_{n2}$ and so on. 

  Consider the entries indexed by the indices $(x_{11}^{(d+1)}, x_{11}^{(d)}), (x_{11}^{(d+1)}, x_{21}^{(d)}), \dots, (x_{11}^{(d+1)}, x_{n1}^{(d)})$. They are $s_1, s_2, \dots, s_n$ respectively and they can be expressed as follows. 
  
  \begin{align*}
    s_j &= \sum_{i_1, i_2, \dots, i_{d-2} \in [n]}x_{1i_1}^{(1)}x_{i_1i_2}^{(2)}\dots x_{i_{(d-2)}j}^{(d-1)} \\
  \end{align*}

  For the other rows indexed by the variables $x_{21}^{(d+1)}, x_{31}^{(d+1)}, \dots, x_{n1}^{(d+1)}$, the sequence of potential non-zero entries is the same ($s_1, s_2, \dots, s_n$) but their positions are shifted by a column compared to the previous non-zero row. See the figure for the block $B_{(d+1),d}$ below for a clearer picture.

  \begin{align*}
    \bbordermatrix
    {
      \text{}& x^{(d)}_{11}& x^{(d)}_{12}& \dotsm& x^{(d)}_{1n}& x^{(d)}_{21}& x^{(d)}_{22}& \dotsm& x^{(d)}_{2n}& \dotsm& x^{(d)}_{n1}& x^{(d)}_{n2}& \dotsm& x^{(d)}_{nn}\cr
      x^{(d+1)}_{11}& {s_1}& 0& \dotsm& 0& {s_2}& 0& 0& \dotsm& \dotsm& {s_n}& 0& \dotsm& 0\cr
      \vdots& \dotsm& \dotsm& \dotsm& \dotsm& \dotsm& \dotsm& 0& \dotsm& \dotsm& \dotsm& \dotsm& \dotsm& \dotsm&\cr
      x^{(d+1)}_{21}& 0& {s_1}& \dotsm& 0& 0& {s_2}& \dotsm& \dotsm& \dotsm& 0& {s_n}& \dotsm& \dotsm&\cr
      \vdots& \dotsm& \dotsm& \dotsm& \dotsm& \dotsm& \dotsm& 0& \dotsm& \dotsm& \dotsm& \dotsm& \dotsm& \dotsm&\cr
      \vdots& \dotsm& \dotsm& \dotsm& \dotsm& \dotsm& \dotsm& \dotsm& \dotsm& \dotsm& \dotsm& \dotsm& \dotsm& \dotsm&\cr
      x^{(d+1)}_{n1}& 0& \dotsm& \dotsm& {s_1}& 0& \dotsm& \dotsm& {s_2}& 0& \dotsm& \dotsm& \dotsm& {s_n}&\cr
      \vdots& \dotsm& \dotsm& \dotsm& \dotsm& \dotsm& \dotsm& 0& \dotsm& \dotsm& \dotsm& \dotsm& \dotsm& \dotsm&\cr
    }
  \end{align*}

  $s_1, s_2, \dots, s_n$ are also the entries indexed by the indices $(1,1), (1,2), \dots, (1,n)$ of the matrix obtained after multiplying the first $(d-1)$ matrices. By induction hypothesis, we know that the entries indexed by the indices $(1,2), \dots, (1,n)$ are not all zero at the point $\ex_0$ which is a zero of the polynomial $\IMM_{n,d}(\ex)$.  This also makes the rows indexed by the variables $x_{11}^{(d+1)}, x_{21}^{(d+1)},\dots, x_{n1}^{(d+1)}$ linearly independent. It is important to note that $P_{11}(\ex) = \IMM_{n,d}(\ex)$. 

  Now, let us define a point such that it is a zero of the polynomial $\IMM_{n,(d+1)}(\ex)$. We set $x_{11}^{(d+1)} = 1$ and $x_{21}^{(d+1)}=x_{31}^{(d+1)}=\dots=x_{n1}^{(d+1)}=0$. Inductively fix the variables appearing in $P_{11}(\ex)$ by the values assigned by $\ex_0$ which is a zero of the polynomial $\IMM_{n,d}(\ex)$. We will fix the other variables suitably later. We call the new point $\ex_0$ as well.

  Now, consider the first $d\times d$ blocks of the Hessian matrix $\hess_{\IMM_{n,(d+1)}}(\ex_0)$. It precisely represents the Hessian matrix of $P_{11}(\ex)$ which is also the Hessian matrix of the polynomial $\IMM_{n,d}(\ex)$ at the point $\ex_0$\footnote{This can be easily seen from the setting of the variables $x_{11}^{(d+1)} = 1$ and $x_{21}^{(d+1)} = x_{31}^{(d+1)} =\dots = x_{n1}^{(d+1)}=0$.}. By induction hypothesis, the rank of this minor of $\hess_{\IMM_{n,(d+1)}}(\ex_0)$ is at least $d(n-1)$. The only non-zero entries in the columns indexed by the variable set $\ex^{(d)}$ are indexed by the variables $x_{11}^{(d)}, x_{21}^{(d)}, \dots, x_{n1}^{(d)}$. This is because the other variables of $\ex^{(d)}$ do not appear in $\IMM_{n,d}(\ex)$.  The row in $B_{(d+1)d}$ indexed by $x_{11}^{(d+1)}$ is the only row that interferes with any of the rows of $B_{1d}, B_{2d},\dots,B_{dd}$. The rows indexed by the variables $x_{21}^{(d+1)}, x_{31}^{(d+1)},\dots, x_{n1}^{(d+1)}$ in $B_{(d+1)d}$ are linearly independent of the rows of $B_{1d}, B_{2d},\dots,B_{dd}$. Hence the rank of $\hess_{\IMM_{n,(d+1)}}$ at the point described is $\geq(d+1)(n-1)$. 

  For the purpose of induction, we must verify that the entries indexed by the indices $(1,2), (1,3), \dots, (1,n)$ of the matrix obtained after multiplying the first $d$ matrices are not all zero at $\ex_0$. These entries are the polynomials $P_{12}, P_{13}, \dots, P_{1n}$. We shall express each of the polynomials in terms of $s_1, s_2, \dots, s_n$ as follows. 

  \begin{align*}
    P_{12} &= s_1x_{12}^{(d)}+ s_2x_{22}^{(d)}+ \dots+ s_nx_{n2}^{(d)} \\
    P_{13} &= s_1x_{13}^{(d)}+ s_2x_{23}^{(d)}+ \dots+ s_nx_{n3}^{(d)} \\
    &\vdots\\
    P_{1n} &= s_1x_{1n}^{(d)}+ s_2x_{2n}^{(d)}+ \dots+ s_nx_{nn}^{(d)} \\
  \end{align*}
  By induction hypothesis, we already know that $s_2, s_3, \dots, s_n$ are not all zero at $\ex_0$. Notice that the variables in $\ex^{(d)}\setminus\{x_{11}^{(d)}, x_{21}^{(d)}, \dots, x_{n1}^{(d)} \}$ were never set in the previous steps of induction\footnote{Because they do not appear in the polynomial $P_{11}$.}. Therefore, we can fix these variables suitably such that $P_{12}, P_{13},\dots, P_{1n}$ are not all zero when evaluated at the point $\ex_0$ (in fact, we can make all of them non-zero). It is clear that we construct the point $\ex_0$ in polynomial time. This completes the proof. 
\end{proof}

\bibliographystyle{alpha}
\bibliography{ref}

\end{document}